\documentclass[12pt,letterpaper]{JHEP3}         
\usepackage{epsfig}

\def\be{\begin{eqnarray}}
\def\ee{\end{eqnarray}}

\newcommand\para{\paragraph{}}

\newcommand{\eqn}[1]{(\ref{#1})}

\def\Dslash{\,\,{\raise.15ex\hbox{/}\mkern-12mu D}}
\def\Dbarslash{\,\,{\raise.15ex\hbox{/}\mkern-12mu {\bar D}}}
\def\delslash{\,\,{\raise.15ex\hbox{/}\mkern-9mu \partial}}
\def\delbarslash{\,\,{\raise.15ex\hbox{/}\mkern-9mu {\bar\partial}}}
\def\pslash{\,\,{\raise.15ex\hbox{/}\mkern-9mu p}}
\def\calDslash{\,\,{\raise.15ex\hbox{/}\mkern-12mu {\cal D}}}

\def\lae{\mathrel{\mathop{\smash{\lower .5 ex \hbox{$\stackrel<\sim$}}}}}
\def\lae{\mathrel{\mathop{\smash{\lower .5 ex \hbox{$\stackrel>\sim$}}}}}

\preprint{DAMTP-2015-12, DCPT-15/07}
\title{\Large Magnetothermoelectric Response from Holography}

\author{Mike Blake$^1$, Aristomenis Donos$^2$ and Nakarin Lohitsiri$^3$\\

 $1$. Department of Applied Mathematics and Theoretical Physics, University of Cambridge, UK
\\ 
 $2$. Centre for Particle Theory and Department of Mathematical Sciences, Science Laboratories, South Road, Durham, DH1 3LE, UK
\\
 $3$. Trinity College, University of Cambridge, CB2 1TQ, UK
\\ 
{\ } 
\\
{\tt m.a.blake@damtp.cam.ac.uk, aristomenis.donos@durham.ac.uk, nl313@cam.ac.uk}
}

\abstract{In this note we study the effects of a magnetic field on transport using holographic models with broken translational invariance. We show that, after carefully subtracting off non-trivial magnetisation currents, it is possible to express the DC transport currents of the boundary theory in terms of properties of a black hole horizon. This allows us to obtain simple analytic expressions for the electrical, thermoelectric and heat conductivity tensors. Our results apply to both isotropic and anisotropic models, including holographic Q-lattices and to certain theories where translational invariance is broken by linear sources for axions.}

\begin{document}
\pagestyle{plain} \setcounter{page}{1}
\newcounter{bean}
\baselineskip16pt







\section{Introduction}

\para{} Despite the obvious relevance to understanding the rich phenomenology of strange metals, it remains an extremely challenging task to calculate the thermoelectric response of a strongly interacting theory. One avenue to make progress is to study quantum critical theories, where the linear response coefficients are constrained to take a scaling form as a function of temperature, $T$ \cite{damle, miraculous1, miraculous2, hartnollscaling}.

\para{} However for theories with a net charge density, $\rho$, it is much harder to make progress. Indeed, in order to obtain finite transport coefficients it becomes necessary to introduce some mechanism for dissipating momentum. Rather than being intrinsic properties of a critical theory, the transport coefficients now depend on the details of how translational invariance, for instance, is broken.

\para{} Nevertheless, it is still possible to make progress by working perturbatively in some small parameter. In particular, detailed results for transport coefficients have been derived both within a model of relativistic hydrodynamics (perturbative in $\rho/T^2$) \cite{hydro} and using the memory matrix formalism (perturbative in the strength of momentum dissipation) \cite{sandiego, impure}.

\para{} An alternative approach is provided by holography. The last couple of years has seen a large amount of progress in obtaining analytic expressions for DC transport in holographic models in which momentum conservation is violated in some manner. These techniques, which originated in the study of the electrical conductivity in massive gravity \cite{vegh, univdc, davison}, have subsequently been generalised to lattice models \cite{lattices, Qlattices, aristosdc}, theories in which translational invariance is broken by linear axions \cite{andradewithers}, and to the calculation of thermoelectric \cite{thermo, amoretti1,amoretti2} and Hall conductivities \cite{hall}.

\para{} The key advantage of these techniques over other approaches is that, rather than being valid only in some perturbative regime, it is possible to obtain exact expressions for DC transport. The results are therefore valid at all temperatures (that is, including $T^2 \ll \rho$) and for any strength of momentum dissipation. In particular, obtaining these expressions for electric Hall transport recently allowed a novel mechanism for obtaining an anomalous scaling of the Hall angle to be identified \cite{hall}.   

\para{} In addition to this behaviour of the Hall angle, many further anomalous aspects of the strange metal transport are evident in the effects of a magnetic field on charge and heat transport. The original hydrodynamic approach of \cite{hydro} to magnetotransport was motivated by the large Nernst signals detected by Ong et al \cite{nernstong}. Similarly, unusual scaling laws are found in the thermopower, magnetoresistance and Hall Lorentz ratio \cite{hartnollscaling}. Motivated by these results, in this note we calculate the full set of DC magnetothermoelectric transport coefficients for a large class of holographic models.  

\para{}As is typical in holography, the key to performing these calculations is to identify radially independent quantities in the bulk that can be identified with the boundary currents \cite{il, univdc, aristosdc}. However, we will see that the existence of non-trivial magnetisation currents complicates the usual discussion \cite{cooper}. In order for us to obtain radially independent quantities, it will be necessary to first subtract off the contribution of the magnetisation current. Nevertheless, the end result is that we will still be able to express the DC transport currents, and hence response coefficients, solely in terms of properties of a black hole horizon.  

\para{}In Section~\ref{section2} we present the details of our holographic models and the calculation of the DC magnetothermoelectric transport. In Section~\ref{section3} we close with a brief discussion of the significance of these results in the wider context of other approaches to magnetotransport \cite{hydro, amoretti3}. In order to improve the readability of our discussion we have relegated certain technical details, such as the definition of the energy magnetisation density and the results for anisotropic theories, to several appendices. 

\para{}We note that whilst this manuscript was in preparation a paper calculating the magnetothermoelectric response in holographic models with massive gravitons appeared \cite{amoretti3}. Whilst the lattice and axion models we are considering here are more general, the close connection between these models and massive gravity \cite{lattices, andradewithers} means that our calculations and results for DC transport take a similar form to those in \cite{amoretti3}. 

\section{Thermoelectric Transport in a Magnetic Field}
\label{section2} 
\para{} In this section our goal is to calculate the transport coefficients of simple holographic models in the presence of a magnetic field. In particular, we wish to obtain the thermoelectric linear response of our theories in response to an applied electric field, $\vec{E}$, and thermal gradient, $\vec{\nabla}T$. 
\para{} As has been discussed at length in \cite{hydro, cooper} there are subtleties with defining these quantities in the presence of a quantising magnetic field. In particular, the electric and heat currents receive additional contributions from spatial variations in the local magnetisation. These additional magnetisation currents must be subtracted out of the total current in order to obtain the physical transport currents that couple to external probes. We therefore need to decompose the total electric, $\vec{J}^{(\mathrm{tot})}$, and heat, $\vec{Q}^{(\mathrm{tot})}$ currents as
\begin{eqnarray}
\vec{J}^{(\mathrm{tot})} &=& \vec{J}+ \vec{J}^{(\mathrm{mag})} \nonumber \\
\vec{Q}^{(\mathrm{tot})} &=& \vec{Q} + \vec{Q}^{(\mathrm{mag})}
\end{eqnarray}
The magnetisation currents, $\vec{J}^{(\mathrm{mag})}$ and $\vec{Q}^{(\mathrm{mag})}$ have been studied in detail in \cite{cooper}. There it was shown that, at the level of linear response, the magnetisation currents induced by electric and thermal gradients are given by
\begin{eqnarray}
\vec{J}^{(\mathrm{mag})}_{i} &=& \frac{M}{T} \hat{\epsilon}_{ij} \vec{\nabla}_jT \\
 \vec{Q}^{(\mathrm{mag})}_{i} &=& M \hat{\epsilon}_{ij} \vec{E}_j + \frac{2 (M_E - \mu  M)}{T} \hat{\epsilon}_{ij} \vec{\nabla}_j T
 \label{magcurrents}
 \end{eqnarray}
where $\hat{\epsilon}_{ij}$ is the 2-dimensional antisymmetric tensor (with $\hat{\epsilon}_{yx} = 1$) and $\mu$ is the chemical potential. The other quantities appearing in \eqn{magcurrents} are the magnetisation density $M$ and the energy magnetisation density $M_E$. We will not need a precise definition of the energy magnetisation density in the main text and so relegate the details to Appendix A. 
\para{}The main focus of our attention is the linear response of the physical transport currents $\vec{J}$ and $\vec{Q}$. This defines the electric, $\hat{\sigma}$, electrothermal, $\hat{\alpha}$, and heat, $\hat{\bar{\kappa}}$ \footnote{Note that $\hat{\bar{\kappa}}$ is not the true thermal conductivity, $\hat{\kappa}$, but rather the thermal conductivity in zero electric field. These are related by $\hat{\kappa} = \hat{\bar{\kappa}} - T \hat{\alpha}. \hat{\sigma}^{-1}. \hat{\alpha}$.}, conductivities according to
\begin{eqnarray}
\vec{J} &=& \hat{\sigma} \vec{E} - \hat{\alpha} \vec{\nabla}T  \nonumber \\
\vec{Q} &=& \hat{\alpha} T \vec{E} - \hat{\bar{\kappa}} \vec{\nabla} {T} 
\label{linearresponse}
\end{eqnarray}
In the presence of a magnetic field, each of these conductivities are $2$ by $2$ matrices. For isotropic systems, which will be the main focus of our discussion, these can be decomposed into their symmetric, e.g. $\sigma_{xx}$, and antisymmetric, e.g. $\sigma_{xy}$, parts\footnote{Note that isotropy implies that $\sigma_{yy} = \sigma_{xx}$ and that $\sigma_{xy} = - \sigma_{yx}$.}. This means that the linear response is described by six functions $\sigma_{xx}, \sigma_{xy}, \alpha_{xx}, \alpha_{xy}, \bar{\kappa}_{xx}, \bar{\kappa}_{xy}$. 
\para{}Our goal then, in this section, is to calculate the DC limit of these transport coefficients for a large class of holographic models. Here, we focus on holographic models that break translational invariance but preserve the homogeneity of the bulk action. These models have received a large amount of recent attention \cite{Qlattices, aristosdc, andradewithers, hall, thermo}, following the realisation that is possible to obtain analytic expressions for their DC transport properties \cite{univdc, aristosdc}. For concreteness we will work with the following simple bulk action
\be
S = \int \mathrm{d}^4x \sqrt{-g} \bigg [ R - \frac{1}{2}[( \partial \phi)^2 + \Phi(\phi) ((\partial {\chi_1})^2 + (\partial {\chi_2})^2)] + V_T(\phi) - \frac{Z(\phi)}{4} F^2 \bigg]
\label{action}
\ee
although our considerations can be trivially extend to more general actions. The translational invariance of the 2+1 dimensional boundary theory is broken by constructing background solutions where we have $\chi_1 = k_1 x$ and $\chi_2 = k_2 y$. Nevertheless, since only derivatives of the $\chi_i$ fields feature in the action, the system remains homogeneous and can be studied using ODEs. For simplicity of presentation we will only consider the isotropic case $k_1 = k_2 = k$ in the main text, although we present results for anisotropic systems in Appendix B. 
\para{}The class of models described by \eqn{action} includes many of the theories that have been studied in the literature. In particular, if we choose to set $\phi = 0, \Phi = \mathrm{const}$ then the $\chi$ fields correspond to massless axions in the bulk. These are dual to marginal operators, $O_{\chi}$ in the boundary theory, in which translational invariance is broken by a linear source $\chi^{(0)}_i = k x_i$. The DC transport properties of these theories have been intensely studied and are directly related to those of massive gravity theories \cite{andradewithers}.
\para{} On the other hand it is also possible to obtain an action of the form \eqn{action} starting from the Q-lattice models introduced in \cite{Qlattices}. These Q-lattice models break translational invariance by introducing an oscillatory lattice through two complex scalar fields $\Psi_1 \sim e^{i k x}$, $\Psi_2 \sim e^{i k y}$. The canonical action of these charged scalars can be rewritten in the form \eqn{action}, which is convenient for discussing transport, by performing the polar decomposition\footnote{Note that in the main text we are assuming that the two fields are related by a bulk $Z_2$ symmetry and so can have the same radial profile. We allow for anisotropic configurations in Appendix B.} $\Psi_i(r) = \phi(r) e^{i \chi_{i}(r)}$. The field $\phi(r)$ therefore corresponds to magnitude of these lattices in the bulk, whilst the $\chi_{i}(r)$ can be thought of as their phase. If one uses conventional kinetic terms this results in the action \eqn{action} with the choice $\Phi(\phi) = \phi^2$ and the requirement that we should identify the fields $\chi_i$ under shifts of $\chi_i \rightarrow \chi_i + 2 \pi$. 
\para For isotropic solutions a suitable ansatz for the background metric and gauge field takes the form
\be
ds^2 = - U dt^2 + U^{-1} dr^2 + e^{2V}(dx^2 + dy^2)
\ee
\be
A = a(r) dt - B y dx
\ee
where we assume that the geometry approaches AdS at the boundary $r \rightarrow \infty$. The temporal gauge field $a(r)$ asymptotes to a constant value $\mu$ which is interpreted as the chemical potential of the boundary theory. Likewise $B$ corresponds to the magnetic field in the dual theory. 
\para{}In addition we will assume that there is a regular black hole horizon located at a position $r_+$ in the bulk. Near this horizon we can expand the radially dependent background fields as
\begin{eqnarray}
U &\sim& 4 \pi T (r - r_+) + ... \nonumber \\
a &\sim& a_+ (r - r_+) + ... \nonumber \\
V &\sim& V_+ + ... \nonumber \\
\phi &\sim& \phi_+ + ...
\end{eqnarray}
where $T$ is identified with the temperature of the dual theory. 
\para{}As is standard, the transport coefficients are computed in holography by studying perturbations of the background solution. We will follow the approach introduced in \cite{aristosdc} and calculate the DC conductivity by applying linear sources to the boundary fields. That is we consider the perturbation ansatz
\begin{eqnarray}
A_{x} &=&  - B y + ( -E + \xi a(r)) t + \delta A_x(r) \nonumber \\
A_{y} &=&  \delta A_y(r) \nonumber \\
g_{tx} &=& - \xi t U + e^{2V} \delta h_{tx}(r) \nonumber \\
g_{ty} &=& e^{2 V} \delta h_{ty}(r) \nonumber \\
g_{rx}  &=& e^{2V} \delta h_{rx}(r) \nonumber \\
g_{ry} &=& e^{2 V} \delta h_{ry}(r) \nonumber \\
\chi_1 &=& kx + \delta \chi_1(r) \nonumber \\
\chi_2 &=& ky + \delta \chi_2(r)
\label{ansatz}
\end{eqnarray}
which corresponds to applying an external electric field $E_i = E \delta_{i x}$ and temperature gradient $(\nabla T)_i = \xi \delta_{i x} T$ to the boundary theory. 
\subsection*{Electrical Currents}
\para{}As was first realised in \cite{univdc, lattices, aristosdc}, the key reason it is possible to calculate transport coefficients in these models is that the currents of the boundary theory can be related to radially-independent quantities in the bulk. In particular, the AdS/CFT dictionary tells us that the expectation value of the currents are given by the quantities
\be
\langle \vec{J}^{(\mathrm{tot}){i}} \rangle  = \sqrt{-g} Z(\phi) F^{i r} \;\;\;\; as \;\;\;\;  r \rightarrow \infty
\ee
Usually the fluxes $\sqrt{-g} Z(\phi) F^{i r}$ are independent of radial position, and so can be evaluated anywhere in the bulk. Evaluating these fluxes at the horizon allows the conductivity tensor to be extracted. 
\para{}However, in this case, the presence of non-trivial magnetisation currents complicates the discussion. It is simple to use the linearised Maxwell equation $\partial_{\mu} (\sqrt{-g} Z(\phi) F^{i \mu})=0$ to show that for the perturbations \eqn{ansatz} we have 
\begin{eqnarray}
\partial_{r} (\sqrt{-g} Z(\phi) F^{ x r}) &=& - \partial_{t} (\sqrt{-g} Z(\phi) F^{ x t})  = 0 \nonumber \\
\partial_{r} (\sqrt{-g} Z(\phi) F^{y r} ) &=& -\partial_{t} (\sqrt{-g} Z(\phi) F^{ y t}) =  - e^{-2V} Z(\phi) B \xi
\label{maxwell}
\end{eqnarray}
which implies that the fluxes are no longer constant in the presence of a thermal gradient. Note that if we had also applied a thermal gradient in the $y$ direction, we would also have found that $\sqrt{-g} Z(\phi) F^{xr}$ depended on the radial coordinate, $r$.
\para{} Nevertheless, we can still construct quantities that are independent of the radial coordinate by integrating equations \eqn{maxwell}. That is we define bulk fluxes by 

%
\begin{eqnarray}
{\cal J}^{x}(r) &=& \sqrt{-g}Z(\phi) F^{x r}  \nonumber \\
{\cal J}^y (r) &=& \sqrt{-g} Z(\phi) F^{y r} - \xi M(r) 
\label{chargecurrent}
\end{eqnarray}
where $M(r)$ is defined to be
\be
M(r) = - \int_{r_+}^{r} d \tilde{r} e^{-2 V} Z(\phi) B
\label{magdensity}
\ee
This extra term has been chosen so that the currents ${\cal J}^{i}$ defined in \eqn{chargecurrent} are radially constant by construction, 
\begin{eqnarray}
\partial_{r} {\cal J}^{i} = 0 
\end{eqnarray}  
However, as we approach the boundary, $r \rightarrow \infty$ we can no longer identify them with the total currents in the boundary theory. Nevertheless, to calculate the response coefficients it is the transport currents, rather than the total currents, that we are interested in. Remarkably, we show in Appendix A that as $r \rightarrow \infty$ then $M(r)$ corresponds to the magnetisation density of the boundary theory. The effect of the additional term in \eqn{chargecurrent} is therefore simply to subtract off the magnetisation current so that near the boundary we have
\be
\langle \vec{J}^{i} \rangle = {\cal J}^{i}(r) \;\;\;\; as \;\;\;\;  r \rightarrow \infty
\ee
i.e. these constant bulk fluxes ${\cal J}^{i}$ precisely correspond to the transport currents of the boundary theory\footnote{Note the the reason we have only had to subtract off the magnetisation current from ${\cal J}^y$ is because our ansatz \eqn{ansatz} only corresponds to applying a thermal gradient in the $x$ direction.}. 
\para{}Having related the transport currents to bulk constants we can proceed to calculate the DC transport as normal. Linearising these expressions according to \eqn{ansatz} gives the bulk constants
\begin{eqnarray}
{\cal J}^{x} &=& - Z(\phi) U \delta A_{x}' - Z(\phi) e^{2V} a' \delta h_{tx} -  B Z(\phi) U \delta h_{r y}  \nonumber \\
{\cal J}^{y} &=& - Z(\phi) U \delta A_{y}' - Z(\phi) e^{2V} a' \delta h_{ty} + B  Z(\phi) U \delta h_{r x} - \xi M(r)
\end{eqnarray}
Since these expressions are independent of the radial coordinate, we can choose to evaluate them wherever we like. The trick, as always, is to proceed to the horizon where the constraints of horizon regularity imply that 
\begin{eqnarray}
\delta A_i &=& -\frac{E_i}{4 \pi T} \mathrm{ln}(r - r_+) + O( r - r_+)  \nonumber\\
\delta \chi_i &=& O((r - r_+)^0)  \nonumber \\
\delta h_{ti} &=& U \delta h_{r i} - \frac{\xi_{i} U}{4 \pi e^{2V} T} \mathrm{ln}(r - r_+)  + O(r - r_+)
\label{regularity}
\end{eqnarray}
Note that for our ansatz we have $E_i = \delta_{ix} E$ and $\xi_i = \delta_{ix} \xi$ but we have left the regularity conditions in their general form. 
\para{}Furthermore the definition of $M(r)$ implies that it vanishes at the horizon. We therefore find that that the transport currents can be expressed solely in terms of properties of the horizon\footnote{This should be contrasted with the total current which, since it depends on the magnetisation, is sensitive to the full geometry.}. In particular we have that 
\begin{eqnarray}
{\cal J}^x &=& Z(\phi) E_x  - e^{2V} Z(\phi) a' \delta h_{tx} - Z(\phi)B \delta h_{t y} \bigg|_{r_+} \nonumber \\
{\cal J}^y &=& Z(\phi) E_y - e^{2V} Z(\phi) a' \delta h_{ty} + Z(\phi)B \delta h_{t x} \bigg|_{r_+} 
\label{horizoncurrents}
\end{eqnarray}
All that remains is to determine the values of the graviton fluctuations $\delta h_{ti}$ at the horizon. This can be done by examining the $t-x$ and $t-y$ components of the linearised Einstein equations which read
\begin{eqnarray}
U (e^{4V} \delta h_{tx}')' - (B^2 Z + e^{2V}k^2 \Phi ) \delta h_{tx} + B Z U e^{2V} a' \delta h_{ry} &=& - e^{2V}Z U a' \delta a_x'  \nonumber \\
U (e^{4V} \delta h_{ty}')' -(B^2 Z +  e^{2V} k^2 \Phi ) \delta h_{ty} - B Z U e^{2V} a' \delta h_{rx} &=&- e^{2V} Z U a'  \delta a_y' +  B Z (- E + \xi a(r)) \nonumber \\
\label{einstein}
\end{eqnarray}
After imposing the regularity conditions \eqn{regularity} these reduce to requiring that we satisfy
\begin{eqnarray}
(B^2Z(\phi) + e^{2V}k^2 \Phi(\phi)) \delta h_{t x} -  B Z(\phi) e^{2V} a' \delta h_{ty} &=& - e^{2V} Z(\phi) a' E + e^{2 V} U' \xi  \nonumber \\
(B^2Z(\phi) + e^{2V}k^2 \Phi(\phi)) \delta h_{t y} + B Z(\phi) e^{2V} a' \delta h_{tx} &=& B Z(\phi) E
\label{gravitonhorizon}
\end{eqnarray}
at the horizon. It is then straightforward to invert these equations and substitute for $\delta h_{ti}$ into \eqn{horizoncurrents}. The resulting expressions for  the electrical currents can be compared to \eqn{linearresponse}, which allows us to read off the electrical conductivity tensor as
\begin{eqnarray}
\sigma_{xx} &=&\frac{e^{2 V} k^2 \Phi (\rho^2 + B^2 Z^2 + Z e^{2V} k^2 \Phi)}{B^2 \rho^2 + (B^2 Z + e^{2V} k^2 \Phi)^2} \bigg|_{r_+} \nonumber \\
\sigma_{xy} &=& B \rho \frac{(\rho^2 + B^2 Z^2 + 2 Z e^{2V} k^2 \Phi)}{B^2 \rho^2 + (B^2 Z + e^{2V} k^2 \Phi)^2} \bigg|_{r_+}
\end{eqnarray}
whilst the electrothermal conductivities are
\begin{eqnarray}
\alpha_{xx} &=&\frac{s \rho e^{2V} k^2 \Phi} {B^2 \rho^2 + (B^2 Z + e^{2V} k^2 \Phi)^2} \bigg|_{r_+} \nonumber \\
\alpha_{xy} &=& s B \frac{(\rho^2 + B^2 Z^2 + Z e^{2V} k^2 \Phi)}{B^2 \rho^2 + (B^2 Z + e^{2V} k^2 \Phi)^2} \bigg|_{r_+}
\label{thermo1}
\end{eqnarray}
which we have expressed in terms of the boundary charge density $\rho  = -Z e^{2V} a'$ and entropy density $s = 4 \pi e^{2V}|_{r_+}$. 
\subsection*{Heat currents}
\para{}Up until now we have focused solely on the electrical current. In order to extract the heat conductivity, we need to consider the heat currents of the boundary theory. In \cite{thermo} it was shown that this can be done by considering the bulk two-form $G^{\mu \nu}$ defined by
\begin{equation}
G^{\mu \nu} = 2 \nabla^{\mu} k^{\nu} + Z(\phi) k^{[\mu} F^{\nu]\sigma}A_{\sigma} + \frac{1}{2} (2 a(r) + E x) Z(\phi) F^{\mu \nu}
\label{twoform}
\end{equation}
where $k^{\mu}$ is the vector field $\partial_{t}$. The heat currents can then be identified with this two form in a similar way to how the electrical current is related to the field strength. In particular, at the linearised level we can make the identification 
\be
\langle \vec{Q}^{(\mathrm{tot}){i}} \rangle  =  \sqrt{-g} G^{r i} \;\;\;\; as \;\;\;\;  r \rightarrow \infty
\ee
which follows from evaluating $G^{r i}$ for the perturbations \eqn{ansatz} to get
\be
\langle \vec{Q}^{(\mathrm{tot}) i} \rangle = U^2 \bigg( \frac{e^{2V} \delta h_{ti}}{U} \bigg)' - a(r)\sqrt{-g}Z(\phi) F^{ir}  \;\;\; as \;\;\; r \rightarrow \infty
\label{heatexp}
\ee
Up to contact terms, the first term in this expression corresponds to the expectation value of the energy momentum tensor $\langle T^{(\mathrm{tot})i0} \rangle$ \cite{thermo}. The second term subtracts off the electric current to get the heat current $\vec{Q}^{(\mathrm{tot}) i} = T^{\mathrm{(tot)}i0} - \mu \vec{J}^{(\mathrm{tot})i}$. 
 \paragraph{}Once we have these fluxes, much of our earlier discussion can now be applied to the heat currents. The motivation for introducing the two-form $G^{\mu \nu}$ is that, in the absence of a thermal gradient, it was shown in \cite{thermo} to satisfy $\partial_{\mu} (\sqrt{-g} G^{\mu i}) = 0$. As a result the linearised fluxes $\sqrt{-g} G^{r i}$ were independent of the bulk radial coordinate. However, the existence of magnetisation currents means that this is no longer true for the perturbations \eqn{ansatz}. Rather we have that
 \begin{eqnarray}
 \partial_{r} ( \sqrt{-g} G^{r x} ) &=& - \partial_{t} (\sqrt{-g} G^{t x} ) - \partial_{y} (\sqrt{-g} G^{y x} ) \nonumber  \\
 &=& 0 \nonumber \\
 \partial_{r} ( \sqrt{-g} G^{r y} ) &=& - \partial_{t} (\sqrt{-g} G^{t y} )-  \partial_{x} (\sqrt{-g} G^{x y} )  +  e^{-2V} Z(\phi) B \xi a(r) \nonumber  \\
 &=&-e^{-2V} Z(\phi) B (E - 2 \xi a(r)) \nonumber
 \ee
 \paragraph{}It is therefore again necessary to add an extra term to the fluxes in order to obtain radially independent constants. That is we construct  
\begin{eqnarray}
{\cal Q}^{x} &=& U^2 \bigg (\frac{e^{2V} \delta h_{tx}}{U} \bigg)' - a(r)  \sqrt{-g} Z(\phi) F^{x r} \nonumber \\
{\cal Q}^{y} &=& U^2 \bigg (\frac{e^{2V} \delta h_{ty}}{U} \bigg)' - a(r)  \sqrt{-g} Z(\phi) F^{y r} - M(r) E  - 2 M_Q(r) \xi
\label{heatcurrent}
\end{eqnarray}
where $M_Q(r)$ is given by 
\be
M_Q (r) =  \int_{r_+}^{r} d \tilde{r} e^{-2 V} Z(\phi) B a(\tilde{r})
\label{heatdensity}
\ee
The additional terms in the definition of ${\cal Q}^{y}$ ensure that these modified fluxes are radially constant
\be
\partial_{r} {\cal Q}^{i} = 0
\ee
by construction. Once again the fact we have been forced to introduce this extra term reflects the presence of magnetisation currents in the boundary. In Appendix A, we show that as $r \rightarrow \infty$ then $M_Q(r)$ precisely approaches the heat magnetisation density, $M_Q = M_E - \mu M$, of the dual theory. The effect of this additional term is therefore to subtract off the contribution of the magnetisation current from \eqn{heatexp}. More precisely, we have that near the boundary
\be
\langle \vec{Q}^{i} \rangle = {\cal Q}^{i} \;\;\; as \;\; r \rightarrow \infty
\ee
i.e. the bulk constants ${\cal Q}^{i}$ correspond to the heat transport currents of the boundary theory.
\para{} We can now repeat the trick we used with the electrical currents and evaluate these constants at the horizon. The definitions of $M_Q(r)$ and $M(r)$ imply that they vanish at $r=r_{+}$ and so, as for the electrical case, we see that the transport currents can be expressed locally in terms of horizon fields. Using the regularity conditions \eqn{regularity} we find that these expressions take the simple form
\begin{eqnarray}
{\cal Q}^{x} &=& - U' e^{2V} \delta h_{tx} |_{r_+}  \nonumber \\
{\cal Q}^{y} &=& - U' e^{2V} \delta h_{ty} |_{r_+}
\end{eqnarray}
\para{}Fortunately, in our discussion of the electrical current we have already determined the values of $\delta h_{t i}|_{r_+}$ by inverting \eqn{gravitonhorizon}. We can therefore extract the thermoelectric conductivity, ${\alpha}$, as
\begin{eqnarray}
{\alpha}_{xx} &=&\frac{s \rho e^{2V} k^2 \Phi} {B^2 \rho^2 + (B^2 Z + e^{2V} k^2 \Phi)^2} \bigg|_{r_+} \nonumber \\
{\alpha}_{xy} &=& s B \frac{(\rho^2 + B^2 Z^2 + Z e^{2V} k^2 \Phi)}{B^2 \rho^2 + (B^2 Z + e^{2V} k^2 \Phi)^2} \bigg|_{r_+}
\end{eqnarray}
which reassuringly agrees with the expression we obtained from the electrical current \eqn{thermo1}. Finally, the heat conductivity, $\bar{\kappa}$ reads
\begin{eqnarray}
\bar{\kappa}_{xx} &=&\frac{ s^2 T (B^2 Z + e^{2V} k^2 \Phi)}  {B^2 \rho^2 + (B^2 Z + e^{2V} k^2 \Phi)^2} \bigg|_{r_+} \nonumber \\
\bar{\kappa}_{xy} &=& \frac{ s^2 T \rho B }{B^2 \rho^2 + (B^2 Z + e^{2V} k^2 \Phi)^2} \bigg|_{r_+}
\end{eqnarray}

\section{Discussion}
\label{section3}

\para{} Whilst on first glance the expressions for these transport coefficients seem rather baroque, on closer inspection they display a remarkable simplicity. Perhaps the most striking aspect of the equations is that, just as in the hydrodynamic analysis of \cite{hydro}, the entire set of DC transport coefficients are described by two parameters. That is, aside from thermodynamic factors, it is only the functions\footnote{Here ${\cal E}$ is the energy density, ${\cal P}$ the pressure and we have identified $\sigma_{ccs}$ as a `charge-conjugation symmetric' conductivity.}
\be
\sigma_{ccs} = Z(\phi)|_{r_+} \;\;\;\;\; \frac{ {\cal E + P}}{\tau} = e^{2V} k^2 \Phi|_{r_+}
\ee
that appear in the transport coefficients. Here we have defined the timescale $\tau$ so that with the above identifications the electric conductivity tensor takes precisely the same form
as in \cite{hydro}, where $\tau^{-1}$ corresponded to the momentum dissipation rate.
\para{}Although the agreement of the electrical conductivity tensor between these two approaches is striking, it does not extend to the
thermoelectric response coefficients (as has previously been emphasised in \cite{amoretti1, amoretti2, amoretti3}). This is not necessarily a surprise since, as we stressed in the introduction, as soon as one includes a net charge density then the transport coefficients will depend on the microscopic way in which momentum dissipation is incorporated. The holographic results we have obtained therefore suggest that the mechanism for momentum dissipation in holography is different from the particular model studied in \cite{hydro}. 
\para{}Whilst our results differ from those presented in \cite{hydro}, they take the same qualitative form as the recent results using massive gravity \cite{amoretti3}. In particular, we reproduce the results of linear axions (and hence massive gravity \cite{andradewithers}) by choosing to take the functions $Z(\phi)$ and $\Phi(\phi)$ to be constant. Although the structure of the equations is very similar, we emphasise that our results apply for much more general holographic models - regardless of the form of these functions. 
\para{}In particular, by varying the choice of action, it is now possible to obtain quite general scalings (with temperature) in the horizon quantities $Z(\phi)|_{r_+}$ and $\Phi(\phi)|_{r_+}$. It is therefore tempting to see if one can choose these scalings to match the phenomenology of the cuprates. This idea was anticipated in \cite{amoretti3} where, after matching $\tau$ and $\sigma_{ccs}$ to the Hall angle as proposed in $\cite{hall}$, the scalings of the thermoelectric and heat transport coefficients, to leading order in $B$, could be deduced. However, since the holographic results differ in general from the hydrodynamic analysis, it is not yet clear to what extent these scalings are universal and hence can be meaningfully compared to experiment.  
\para{}Nevertheless, the models and results we have presented here should be capable of realising these proposed scalings explicitly. Furthermore, since our results are valid even in strong magnetic fields, it is possible to use these models to go beyond leading order in $B$. In particular, it has recently been proposed that the magnetoresistance of strange metals can be attributed to the effects of the magnetic field on the critical theory itself \cite{magnetoresistance}. It would therefore be extremely interesting to use the holographic models discussed here to study transport in the presence of a strong magnetic field.

\section*{Acknowledgements}

We're grateful to Koenraad Schalm, David Tong and Jan Zaanen for useful discussions. We also thank Jerome Gauntlett for spotting several typos in the first version of this paper. This work was supported by the European Research Council under the European Union's Seventh Framework Programme (FP7/2007-2013), ERC Grant agreement STG 279943, Strongly Coupled Systems. MB is funded by a Junior Research Fellowship at Churchill College, Cambridge.

\appendix

\section{Magnetisation and Energy Magnetisation Densities}

In this Appendix we wish to derive the formulae for the magnetisation and energy magnetisation densities used in the main text. The definition of the magnetisation is familiar. If we apply a magnetic field to the boundary theory via a source $A_x^{(0)} = - B y$, then the magnetisation density is given by differentiating the Euclidean action $S_E$, as
\be
M = -\frac{1}{{\cal V}} \frac{\partial{S_E}}{\partial{B}}
\ee
where ${\cal V}$ is the volume of the boundary field theory. 
\para{}The energy magnetisation density is defined as an analogous quantity for the metric. That is we should apply a source $\delta g_{tx}^{(0)} = -B_1 y$ and differentiate with respect to $B_1$
\be
M_E = -\frac{1}{{\cal V}} \frac{\partial{S_E}}{\partial{B_1}}\bigg |_{B_1 = 0}
\ee
We now wish to calculate these for the background solutions to our action \eqn{action}. To do this, it is convenient to consider solutions obeying the ansatz
\be
\chi_i &=& k x_i \;\;\; \phi = \phi(r) \nonumber 
\ee
\be
A_t &=& a(r) \;\;\;\;\;\; A_x = - B y + (a(r) - \mu) B_1 y  \nonumber 
\ee
\be
ds^2 &=& -U(r) (dt + B_1 y dx)^2 + \frac{dr^2}{U(r)} + e^{2 V(r)} (dx^2 + dy^2) \nonumber 
\label{kaluza}
\ee
When $B_1 = 0$, these are simply the background solutions we studied in the main text. At leading order in $B_1$ we also have applied a source $\delta g_{tx}^{0} = - B_1 y$ to the boundary theory which will allow us to evaluate the energy magnetisation density. The higher order terms in $B_1$ in the metric ensure that we have written down a consistent ansatz. In particular note that even though we have introduced dependence on the $y$-coordinate, this does not appear in the equations of motion. The dynamical fields $\phi(r), a(r), U(r), V(r)$ only depend on the radial coordinate. 
\paragraph{}In order to calculate the magnetisation and energy magnetisation densities, we need to differentiate the action with respect to $B$ and $B_1$, before setting $B_1$ to zero. To do this we first differentiate the off-shell bulk action, before evaluating these derivatives on the equations of motion. With our ansatz we find that the Einstein-Hilbert term in the action \eqn{action} can be written as
\be
\frac{1}{{\cal V}} S^{\mathrm{EH}} = \int_{r_+}^{\infty} \mathrm{d}r \frac{e^{-2 V}}{2}\bigg[ U B_1^2 - 2 e^{4V}(U'' + 4 U' V' + 6 U V'^2 + 4 U V'') \bigg] \nonumber
\ee
Similarly the scalar terms take the form
\be
\frac{1}{{\cal V}} S^{\mathrm{scalar}} = - \int_{r_+}^{\infty} \mathrm{d}r \bigg[ \frac{1}{2} e^{2V} U \phi'^2 +  k^2 \Phi(\phi) - e^{2V} V_T (\phi) \bigg] \nonumber
\ee
Finally we have the Maxwell term 
\be
\frac{1}{{\cal V}} S^{\mathrm{Maxwell}} = - \int_{r_+}^{\infty} \mathrm{d}r \frac{e^{-2V}}{2} Z(\phi) \bigg[ (B + B_1 \mu)^2 - 2 B_1 (B + B_1 \mu) a + B_1^2 a^2 - e^{4V} a'^2 \bigg]  \nonumber
\ee
Although we will not need them here, the equations of motion can of course be deduced by varying this action with respect to the dynamical fields $a(r), \phi(r), U(r), V(r)$.
\para{}To get the magnetisation density, we simply need to set $B_1=0$ and then differentiate the action. We note that the only explicit $B$ dependence is in the Maxwell term and hence we find\footnote{Note that the Euclidean action constructed via a Wick rotation $t \rightarrow - i \tau$ has an extra minus sign relative to the Lorentzian action \eqn{action}.}
\be
M = -\frac{1}{{\cal V}} \frac{\partial S_E}{\partial B} = - \int_{r_+}^{\infty} \mathrm{d}r  e^{-2V} Z(\phi) B
\ee
It is now clear to see that the $M(r)$ defined in \eqn{magdensity} is precisely the magnetisation density when $r \rightarrow \infty$. 
\para{} Likewise we can construct the energy magnetisation density by first differentiating with respect to $B_1$ and then setting $B_1 = 0$. At linear order in $B_1$ we again find that it is only the Maxwell term that contributes\footnote{Note that at leading order in $B_1$ the magnetisation and energy magnetisation densities do not receive any contribution from the boundary counterterms.}. We thus read off the energy magnetisation density as 
 \be
M_E = -\frac{1}{{\cal V}} \frac{\partial S_E}{\partial B_1}\bigg|_{B_1= 0} = -\int_{r_+}^{\infty} \mathrm{d}r e^{-2V} Z(\phi) B (\mu  - a(r))
 \ee
Finally we deduce that the heat magnetisation density is given by
 \be
 M_Q = M_E - \mu M = \int_{r_+}^{\infty} \mathrm{d}  e^{-2V} Z(\phi) B a(r)
 \ee
 from which we can see that the function $M_Q(r)$ \eqn{heatdensity} is equivalent to $M_Q$ when $r \rightarrow \infty$.
 
 \section{Anisotropic Models}

It is straightforward to generalise our calculations to anisotropic theories. In particular, we can consider the action
\be
S = \int \mathrm{d}^4x \sqrt{-g} \bigg [ R - \frac{1}{2}[( \partial \phi)^2 + \Phi_1(\phi) (\partial {\chi_1})^2 + \Phi_2(\phi) (\partial {\chi_2})^2] + V_T(\phi) - \frac{Z(\phi)}{4} F^2 \bigg] \nonumber
\label{anisotropic}
\ee
where we now break translational invariance in the $x$ and $y$ directions by constructing background solutions with $\chi_1 = k_1 x$ and $\chi_2 = k_2 y$. In order to allow for anisotropic solutions we should also modify our metric ansatz to
\be
ds^2 = - U dt^2 + U^{-1} dr^2 + e^{2V_1}dx^2 + e^{2V_2} dy^2
\ee
The resulting expressions for DC transport are more complicated that the isotropic case, but are again simplified somewhat by introducing the thermodynamic factors $s = 4 \pi e^{V_1 + V_2}|_{r_+}$ and $\rho = -e^{V_1 + V_2} Z(\phi) a'$. They read
\be
\sigma_{xx} = \frac{e^{V_1 + V_2} k_2^2 \Phi_2 (\rho^2 + B^2 Z^2 + Z e^{2V_2} k_1^2 \Phi_1 ) }{B^2 \rho^2 + (B^2 Z + e^{2V_2} k_1^2 \Phi_1)(B^2 Z + e^{2V_1} k_2^2 \Phi_2 )} \bigg|_{r_+}
\ee
\be
\sigma_{xy} = B \rho \frac{(\rho^2 + B^2 Z^2 + Z e^{2V_2} k_1^2 \Phi_1 + Z e^{2V_1} k_2^2 \Phi_2) }{B^2 \rho^2 + (B^2 Z + e^{2V_2} k_1^2 \Phi_1)(B^2 Z + e^{2V_1} k_2^2 \Phi_2 )} \bigg|_{r_+}
\ee
\be
\alpha_{xx} = \frac{ s \rho e^{V_1 + V_2} k_2^2 \Phi_2}{B^2 \rho^2 + (B^2 Z + e^{2V_2} k_1^2 \Phi_1)(B^2 Z + e^{2V_1} k_2^2 \Phi_2 )} \bigg|_{r_+}
\ee
\be
\alpha_{xy} = \frac{s B (\rho^2 + B^2 Z^2 + Z e^{2V_2} k_1^2 \Phi_1 ) }{B^2 \rho^2 + (B^2 Z + e^{2V_2} k_1^2 \Phi_1)(B^2 Z + e^{2V_1} k_2^2 \Phi_2 )} \bigg|_{r_+}
\ee
\be
\bar{\kappa}_{xx} = \frac{4 \pi e^{2 V_2} s T (B^2 Z + e^{2V_1} k_2^2 \Phi_2)}{B^2 \rho^2 + (B^2 Z + e^{2V_2} k_1^2 \Phi_1)(B^2 Z + e^{2V_1} k_2^2 \Phi_2 )} \bigg|_{r_+}
\ee
\be
\bar{\kappa}_{xy} = \frac{s^2 T \rho B}{B^2 \rho^2 + (B^2 Z + e^{2V_2} k_1^2 \Phi_1)(B^2 Z + e^{2V_1} k_2^2 \Phi_2 )} \bigg|_{r_+}
\ee
where the expressions for the remaining transport coefficients (e.g. $\sigma_{yy}$) can be trivially obtained from those we have presented above through swapping around the labels 1 and 2.


\begin{thebibliography}{99}

 \bibitem{damle} K.~Damle and S.~Sachdev,
  ``{\it Nonzero-temperature transport near quantum critical points},''
  Phys. Rev. B {\bf 56} 8714 (1997),
  arXiv:cond-mat/9705206 [cond-mat.str-el].
  
\bibitem{miraculous1}
M. J. Bhaseen, A. G. Green, S. L. Sondhi,
``Magnetothermoelectric Response at a Superfluid--Mott Insulator Transition'',
Phys.\ Rev.\ Lett {\bf 98 } 166801 (2007)
	arXiv:cond-mat/0610687 [cond-mat.str-el]

\bibitem{miraculous2}
M. J. Bhaseen, A. G. Green, S. L. Sondhi,
``Magnetothermoelectric Response near Quantum Critical Points'',
Phys.\ Rev.\ B {\bf 79 } 094502 (2009)
arXiv:0811.0269 [cond-mat.str-el]

\bibitem{hartnollscaling}
  S.~A.~Hartnoll and A.~Karch,
  ``Scaling theory of the cuprate strange metals,''
  arXiv:1501.03165 [cond-mat.str-el].

\bibitem{hydro} S.~A.~Hartnoll, P.~K.~Kovtun, M.~Muller and S.~Sachdev,
  ``{\it Theory of the Nernst effect near quantum phase transitions in condensed matter, and in dyonic black holes},''
  Phys.\ Rev.\ B {\bf 76}, 144502 (2007)
  [arXiv:0706.3215 [cond-mat.str-el]].

\bibitem{sandiego} S.~A.~Hartnoll and D.~M.~Hofman,
  ``{\it Locally Critical Resistivities from Umklapp Scattering},''
  Phys.\ Rev.\ Lett.\  {\bf 108}, 241601 (2012)
  [arXiv:1201.3917 [hep-th]].
  
\bibitem{impure}
  S.~A.~Hartnoll and C.~P.~Herzog,
  ``Impure AdS/CFT correspondence,''
  Phys.\ Rev.\ D {\bf 77} (2008) 106009
  [arXiv:0801.1693 [hep-th]].

 \bibitem{vegh}   D.~Vegh,
  ``{\it Holography without translational symmetry},''
  arXiv:1301.0537 [hep-th].

 \bibitem{univdc} M.~Blake and D.~Tong,
``{\it Universal Resistivity from Holographic Massive Gravity},''
Phys.\ Rev.\ D {\bf 88}, 106004 (2013)
[arXiv:1308.4970 [hep-th]].
 
 \bibitem{davison}
R.~A.~Davison,
``{\it Momentum relaxation in holographic massive gravity},''
[arXiv:1306.5792 [hep-th]].
 
   \bibitem{lattices}   M.~Blake, D.~Tong and D.~Vegh,
  ``{\it Holographic Lattices Give the Graviton a Mass},''
  Phys. Rev. Lett. {\bf 112} 071602 (2014)
  [arXiv:1310.3832 [hep-th]].
    
   \bibitem{Qlattices} A.~Donos and J.~Gauntlett,
  ``{\it Holographic Q-lattices},''
  JHEP {\bf 1404}, 040 (2014) 
 [ arXiv:1311.3292 [hep-th]]
 
 \bibitem{aristosdc} A.~Donos and J.~Gauntlett,
 ``{\it Novel metals and insulators from holography},''
 [arXiv:1401.5077 [hep-th]]
    
\bibitem{andradewithers}
  T.~Andrade and B.~Withers,
  ``A simple holographic model of momentum relaxation,''
  JHEP {\bf 1405} (2014) 101
  [arXiv:1311.5157 [hep-th]].
  
\bibitem{thermo}
  A.~Donos and J.~P.~Gauntlett,
  ``Thermoelectric DC conductivities from black hole horizons,''
  JHEP {\bf 1411} (2014) 081
  [arXiv:1406.4742 [hep-th]].
  
\bibitem{amoretti1}
  A.~Amoretti, A.~Braggio, N.~Maggiore, N.~Magnoli and D.~Musso,
  ``Thermo-electric transport in gauge/gravity models with momentum dissipation,''
  JHEP {\bf 1409} (2014) 160
  [arXiv:1406.4134 [hep-th]].
  
\bibitem{amoretti2}
  A.~Amoretti, A.~Braggio, N.~Maggiore, N.~Magnoli and D.~Musso,
  ``Analytic dc thermoelectric conductivities in holography with massive gravitons,''
  Phys.\ Rev.\ D {\bf 91} (2015) 2,  025002
  [arXiv:1407.0306 [hep-th]].
  
\bibitem{hall}
  M.~Blake and A.~Donos,
  ``Quantum Critical Transport and the Hall Angle in Holographic Models,''
  Phys.\ Rev.\ Lett.\  {\bf 114} (2015) 021601
  [arXiv:1406.1659 [hep-th]].
  
  \bibitem{nernstong} Z.~A.~Xu, N. P~.~ Ong, Y.~Wang, T.~Kakeshita, and S.~Uchida,
Nature 406, 486 (2000).
  
  \bibitem{il}   N.~Iqbal and H.~Liu,
  ``{\it Universality of the hydrodynamic limit in AdS/CFT and the membrane
paradigm},''
  Phys.\ Rev.\ D {\bf 79}, 025023 (2009)
  [arXiv:0809.3808 [hep-th]].

  \bibitem{cooper} N. R. Cooper, B. I. Halperin, and I. M. Ruzin,
  ``{\it Thermoelectric response of an interacting two-dimensional electron gas in a quantizing magnetic field},''
Phys.\ Rev.\ B {\bf 55}, 2344 (1997)
  
\bibitem{amoretti3}
  A.~Amoretti and D.~Musso,
 ``Universal formulae for thermoelectric transport with magnetic field and disorder,''
  arXiv:1502.02631 [hep-th].
  
  \bibitem{magnetoresistance}
  I.~M.~Hayes, N.~P.~Breznay, T.~Helm, P.~Moll,
 ``Magnetoresistance near a quantum critical point,''
  [arXiv:1412.6484v2 [cond-mat.str-el]].
  
\end{thebibliography}
\end{document}